\documentstyle[11pt,nyas]{article}

\begin{document}

\def\ni{\noindent}
\def\th{\thinspace}

\title{INFLUENCE OF THE TACHOCLINE ON SOLAR EVOLUTION}

\author{A. S. Brun\altaffilmark{1} and J.-P. Zahn}
\affil{D\'epartement d'Astrophysique Stellaire et Galactique, Observatoire de Paris, Section Meudon, 92195 Meudon, France}

\altaffiltext{1}{Present address: JILA, University of Colorado, Boulder, CO 80309-0440, USA.}

\begin{abstract}
Recently helioseismic observations have revealed the pre- sence of a shear layer at the base of the convective zone related to the transition from differential rotation in the convection zone to almost uniform rotation in the radiative interior, the tachocline. At present, this layer extends only over a few percent of the solar radius and no definitive explanations have been given for this thiness. Following Spiegel and Zahn (1992, Astron. Astrophys.), who invoke anisotropic turbulence to stop the spread of the tachocline deeper in the radiative zone as the Sun evolves, we give some justifications for their hypothesis by taking into account recent results on rotating shear instability (Richard and Zahn, 1999, Astron. Astrophys.). We study the impact of the macroscopic motions present in this layer on the Sun's structure and evolution by introducing a macroscopic diffusivity $D_T$ in updated solar models. We find that a time dependent treatment of the tachocline significantly improves the agreement between 
computed and observed surface chemical species, such as the $^7$Li and modify the internal structure of the Sun (Brun, Turck-Chi\`eze and Zahn, 1999, Astrophys. J.).\\

{\small \bf
\ni to appear in the Annals of the New York Academy of Sciences, Vol 898.
}

\end{abstract}



\vspace{-0.5cm}

\section{INTRODUCTION}
The presence of a shear layer connecting the differential rotation of the convective zone to the solid rotation of the radiative zone is now well established by helioseismic inversions (Figure 1 and [1]). Both its location and width are more and more constrained and seem to be $0.691\pm0.004$ $R_{\odot}$ and less than 0.05 $R_{\odot}$, respectively. Today, there are several hydrodynamical or MHD descriptions of this shear layer and its extension but none is definitive [2], [3], [4]. In these models the motions in the tachocline are either turbulent or laminar, involve magnetic field or are purely hydrodynamical. In this paper we discuss the basic ideas supporting the dynamical description of this transition layer. First, in Section 2 we recall the physical processes acting in this layer and summarize the different approaches with an emphasis on Spiegel and Zahn's description invoking a nonlinear anisotropic turbulence. In Section 3, using the prescription of Spiegel and Zahn [2] for the amplitude of the vertical velocity in the tachocline and Chaboyer and Zahn [5] for the chemical mixing and evolution, we build solar models including a macroscopic diffusivity $D_T$, which are compared to the most recent helioseismic data and surface abundance observations for $^7$Li, $^9$Be and $^3$He/$^4$He ratio. Finally, we summarize our results and conclude in Section 4.

\section{THE SOLAR TACHOCLINE}

The transition layer between the convective and radiative zones (Figure 1) plays a crucial role in our understanding of stars such as the Sun because it simultaneously involves several physical features, such as:

\begin{itemize}
\item the strong shear associated with the transition from differential to solid rotation, which may generate turbulence,
\item a turbulent interface with the convection zone above, which may produce internal waves [6], [7],
\item the possible presence of magnetic field, of fossile origin [3], or linked with the 11-year cycle [8],
\item  the proximity of the thermonuclear burning zone of lithium 7 and beryllium 9.
\end{itemize}  
 
\begin{figure}[!htb]
\setlength{\unitlength}{1.0cm}
\begin{picture}(9,4.6)
\includegraphics{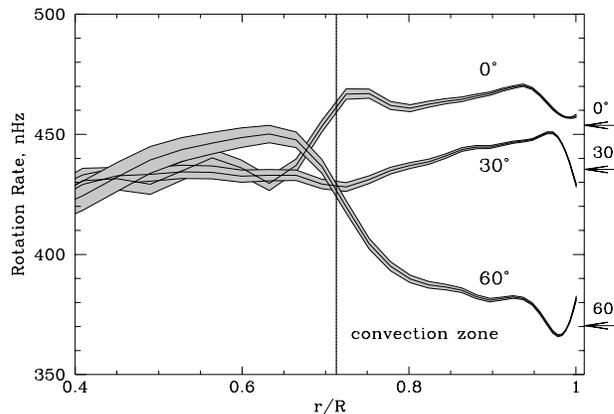}
\end{picture}
\caption[]{\small Solar rotation rate inferred from MDI data (aboard SOHO) as a function of radius at three latitudes, 0$^o$, 30$^o$ and 60$^o$ (the vertical line represents the base of the convection zone) [9]. }
\end{figure}

In this section we shall concentrate on some specific points concerning the 
tachocline and we will not deal with the full complexity of this transition 
layer.\\ 
The presence of a latitudinal differential rotation at the base of the 
convection zone induces a latitudinal temperature gradient  
${\Omega}(\theta)\rightarrow \nabla T(\theta)$.  Without any limitating 
process this temperature gradient will diffuse inwards, on a thermal 
diffusion time scale, enforcing differential rotation deep into the 
radiative interior of the Sun. As we already stated, however, helioseismic 
observations indicate a uniform rotation profile in the radiation zone, and 
thus we have to find which processes could hinder this diffusion.\\
There are different possibilities and first and foremost is the stable 
stratification of the radiation zone. This effect would indeed slow down the 
spread of the tachocline, but in spite of that, the layer
  would still extend to one 
third of the radius in the present Sun, as estimated by Spiegel and Zahn [2], 
which is far too much. 
Thus another process must be invoked to explain the observed thinness of the 
tachocline.\\
One possibility is the presence of a magnetic field in the radiation zone, 
as advocated by Gough and McIntyre [3]. Their model is promising, but it
has not yet been worked 
out in detail: it remains to be seen how the poloidal field threads into 
the convection zone, and avoids imposing differential rotation throughout the 
radiative interior.\\
Another possibility has been suggested by Spiegel and Zahn [2]. If the latitudinal shear 
is unstable, it could generate anisotropic turbulence which
would tend to reduce the differential rotation, and hence prevent the spread 
of the tachocline. It is not clear whether the solar tachocline is linearly 
unstable: when one applies the criterion derived by Watson [10] to an 
angular velocity profile of the type 
$\Omega \propto 1 - \alpha_1 \sin^2 \lambda$, where $\lambda$ is the latitude,
linear instability requires $\alpha_1 > 0.29$, which is larger than the solar value
$\alpha_1 \approx 0.25$. It appears, however, that a law of the type
$\Omega \propto 1 - \alpha_1 \sin^2 \lambda - \alpha_2 \sin^4 \lambda$, closer to the latitudinal dependence drawn from helioseismic inversions (see Fig. 1) would be more sensitive to such instability [11]. Furthermore, a toroidal magnetic field of 
sufficient strength 
would also act to destabilize the flow, as shown by Gilman and Fox [4].
In any case, the Reynolds number is so high that such a differential rotation 
would be liable to finite amplitude
instability, as can be infered from laboratory experiments 
[12].\\
It has been argued by Gough and McIntyre [3] that turbulence does not 
necessarily reduce the shear of differential rotation: in the Earth's 
atmosphere the transport of angular momentum would even imply a negative 
viscosity. But there the transport is achieved mainly through Rossby waves,
and it is not clear whether this can be applied to the solar tachocline.\\
Admittedly, this important issue is still a matter of debate, and it may be 
settled only by comparing the models' predictions with the observed properties 
of the Sun. This is why we have chosen to draw all observable consequences
from the model which has been worked out in sufficient detail to allow
such a test, namely
the turbulent tachocline proposed by Spiegel and Zahn [2].
We will show that the mixing induced in this model 
 improves both the sound speed profile (reduction of 
the peak below the convection zone) and the surface light element abundances, 
confirming the need of introducing macroscopic processes in solar models 
[13]-[15].

\section{MIXING IN THE SOLAR TACHOCLINE: PHYSICAL DESCRIPTION}

Macroscopic mixing may be treated in solar models by adding an effective 
diffusivity $D_T$ in the equation for the time evolution of the concentration 
of 
chemical species. To establish this coefficient for the tachocline, we use the 
description by Spiegel and Zahn [2], where anisotropic turbulence is responsible for stopping the spread of the layer.  This 
anisotropic diffusion will also interfere with the advective transport of 
chemicals. Chaboyer and Zahn [5] have shown that the result is a diffusive 
transport in the vertical direction. 
Using their result, Brun, Turck-Chi\`eze and Zahn [16] derived the following 
expression for the macroscopic diffusivity:
\begin{equation}\label{coef}
D_T(r)=\frac{4}{405} \nu_H \left(\frac{d}{r_{bcz}}\right)^2 \mu_4^6 \, Q_4^2 \exp(-2\zeta) \cos^2(\zeta) + \mbox{higher order terms} 
\end{equation}
with $Q_{4}=\tilde{\Omega}_{4}/\Omega$, $\tilde{\Omega}_{4}$ characterizing the differential rotation rate,  $\mu_4=4.933$, $\zeta=\mu_{4} (r_{bcz}-r)/d$ a non-dimensional depth,  
\begin{equation}\label{d}
d=r_{bcz}(2\Omega/N)^{1/2}(4K/\nu_H)^{1/4}
\end{equation}
 a length related to the tachocline thickness $h$ (e.g $h \sim d/2$), $r_{bcz}$ the radius, $\Omega$ the 
angular velocity and $K=\chi/\rho c_p$ the radiative diffusivity at the base of the convective zone. The horizontal component of the macroscopic diffusivity $D_H$ is assumed to be equal to the horizontal viscosity $\nu_H$. In our solar models, we treat $h$ (hence $d$) as an adjustable parameter, chosen to agree with the helioseismic determination of the tachocline thickness $h\leq0.05 R_{\odot}$ [1]. With the latitudinal dependence of the angular velocity at the base of the convection zone deduced from Thompson et al. [17], $\Omega_{bcz}/2\pi=456-72x^2-42x^4$ nHz (with $x=\sin\lambda$), we have 
reestimated the coefficient $Q_4=-1.707\times10^{-2}$, as well as the ratio between the rotation in the deep radiative zone and the equatorial rate $\Omega/\Omega_0=0.9104$. The prediction by Gough and Sekii [18] for the latter, who consider instead the magnetic stresses, is $\sim 0.96$; presently, the seismic observations suggest a rotational ratio of $0.94\pm0.01$ [1], which is intermediate between these two theoretical estimates. 

 An analysis of the dependence of our $d$ and $D_T$ with the global and 
differential rotation rates yields 
\begin{equation}
D_T \propto \nu_H \left(\frac{d}{r_{bzc}}\right)^2 Q_i^2 \propto \Omega \nu_H^{1/2} (\hat{\Omega}/\Omega)^2,  \qquad  d \propto 
\Omega^{1/2}/\nu_H^{1/4}. \nonumber
\end{equation}  
where we have used equations (\ref{coef}) and (\ref{d}). 
Assuming that the turbulent viscosity is proportional to the differential 
rotation (i.e., $\nu_H\propto\hat{\Omega}$), as suggested by the laboratory
experiments [12], and introducing the dependence of the 
differential rotation on rotation observed by Donahue, Saar and Baliunas [19] 
($\hat{\Omega}\propto\Omega^{0.7\pm0.1}$), we finally obtain the following 
scalings
\begin{equation}
D_T \propto \Omega^{0.75\pm0.25}, \qquad  d \propto \Omega^{(1.3\mp0.1)/4}. 
\nonumber
\end{equation}
We conclude that the tachocline mixing was stronger in the past both because that 
layer was thicker and because the diffusivity was larger. We render the mixing 
in the tachocline time dependent, through $D_T(\Omega(t))$ and $d(\Omega(t))$, 
by using the spin-down law $\Omega \propto t^{-1/2}$ which was deduced by 
Skumanich [20] from the rotation rate of stellar clusters of different ages.

\section{RESULTS}
Starting from the reference model of Brun, Turck-Chi\`eze and Morel [21] 
built  with the CESAM code [22], we introduce for this study the 
nuclear reaction 
$^7$Li(p,$\alpha$)$^4$He proposed by Engstler et al. [23] and the 
coefficient 
$D_T$ (Eq. 1) in the diffusion equation of chemical species, and we 
follow the time 
evolution of the solar model from the pre-main sequence (PMS) until 4.6 Gyr. The results are shown in the Table~1 and Figures 2-5 (see Ref. [15] for a more detailed discussion). We use a tachocline thickness $h$ of 0.05 
or 0.025 $R_{\odot}$, $N$ of 100 or 25 $\mu$Hz and $\Omega$ of 0.415 
$\mu$Hz. Our standard model [21] has a surface abundance for helium of 0.2427 in mass, corresponding to an $^4$He difffusion of 10.8\%. This value of Y$_s$ is a bit too low if we compare with the Basu and  
Antia [24] value for the OPAL equation of state [25], Y$_s=0.249 \pm 0.003$.

\vspace{-0.5cm}
{\scriptsize
\begin{table}[!ht]
\begin{center}
\caption[]{\small Surface abundance variation of $^3$He/$^4$He 
during the last 3 Gyr, 
surface abundances of $^4$He and heavy elements Z, and abundance ratio 
initial/surface for 
$^7$Li and $^9$Be from observations and for solar models at the solar age$^a$}
\end{center}
\vspace{-0.5cm}
\begin{tabular}{p{1.5cm}*{9}{c}}
\hline
 & Obs & Ref & $A$ & $B$ & $A_t$ & $B_t$ & $B_{tz}$ & $C_t$\\
\hline
 $d$ ($r/R_{\odot}^{.}$)& $\leq0.1$ & - & 0.1 & 0.1 & 0.1 & 0.1 & 0.1 & 0.05\\
 $N$ ($\mu$Hz) & -&- & 100 & 25 & 100 & 25 & 25 & 25 \\
 ($^3$He/$^4$He)$_s$ & max 10\% & 2.28\% & 2.14\% & 2.01\% & 2.11\% & 2.0\% & 
2.02\%  & 2.07\% \\
 $^4$He$_s$ & 0.249$\pm$0.003 &  0.2427 & 0.2452 & 0.2473 & 0.2455 & 0.2477 & 
0.2509  & 0.2464  \\
 (Z/X)$_s$ & 0.0245$\pm$0.002& 0.0245  & 0.0245 & 0.0245 & 0.0245 & 0.0245 & 
0.0255 & 0.0245  \\
 $^7$Li$_0$/$^7$Li$_s$ & $\sim$100 & $\sim$6 & $\sim$8 &$\sim$22 & $\sim$12 
&$\sim$91 &$\sim$134  & $\sim$89 \\
 $^9$Be$_0$/$^9$Be$_s$ & 1.10$\pm$0.03 & 1.115  & 1.093 & 1.086 & 1.093 & 1.118 & 1.125 & 1.092 \\
 \hline
\end{tabular}
$^a$Subscripts $t$ for time dependent models, $tz$ for time dependent model with Z$_0=$Z$_0^{ref}=0.01959$
\end{table}} 

\begin{figure}[htb*]
\setlength{\unitlength}{1.0cm}
\begin{picture}(9,5)
\includegraphics{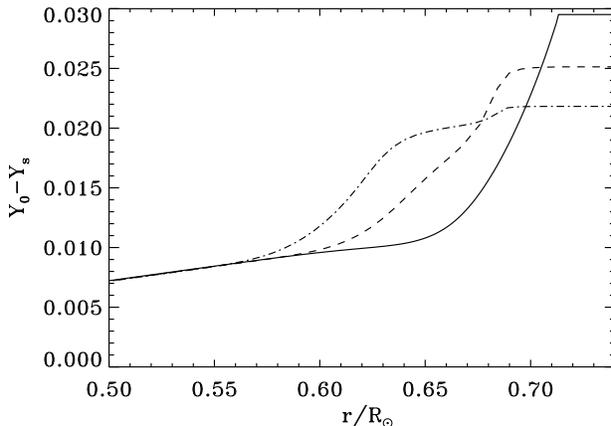}
\end{picture}      
 \caption{\small Radial profile of the difference of  $^4$He composition 
 between the initial and present values for the reference solar model including
  only microscopic diffusion ({\it solid line}) and solar models where we add a  macroscopic mixing due to the presence of the tachocline: coefficient $A$ 
  ({\it dash}) and $B$ ({\it dash dot}) (see Table 1 for the model 
characteristics).}
\end{figure}

\begin{figure}[htb*]
\setlength{\unitlength}{1.0cm}
\begin{picture}(8,5)
\includegraphics{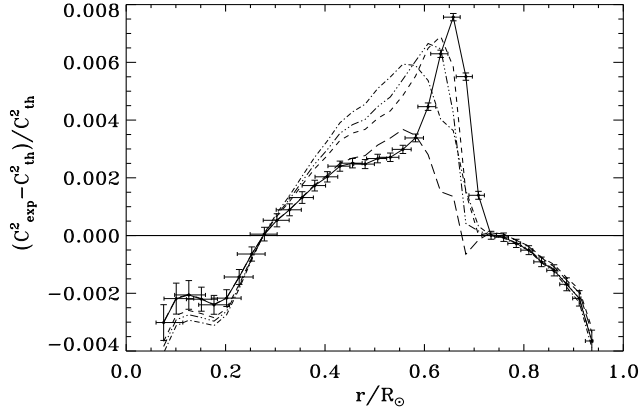}
\end{picture}      
\caption{\small Squared sound  speed difference between GOLF+MDI 
data and the reference model ({\it solid line}) or models including a turbulent term: coef $A_t$ ($d=0.1$, $N=100$) ({\it dash}), $B_t$ ($d=0.1$, $N=25$) ({\it dash dot}), $C_t$ ($d=0.05$, $N=25$) ({\it dash three dots}), and one model with $B_t$ and $Z_{0}=Z_{0}^{ref}=0.01959$ (i.e., $B_{tz}$) ({\it long dash}). }
\end{figure}

 When introducing our diffusive coefficient, we mix helium back into the 
convection zone, inhibiting the microscopic diffusion up to 25\% and producing 
a photospheric $^4$He$_s=0.2473$ 
(cf. Table 1, models $A$ and $B$). As expected, the composition profile is smoother and flattens over the distance $h$ below the convective zone (see Figure 2). 

The effect on the sound speed is displayed in Figure 3. When the macroscopic 
transport is neglected, the squared
sound speed difference reveals a peak just below the convection zone, 
coinciding with the tachocline ({\it solid line}). 
Macroscopic diffusion acts to reduce this peak, but when one recalibrates
the model to yield, e.g., the present abundance of heavy elements
($Z/X=0.0245\pm0.002$), 
the effect is rather minor. On the other hand, if the heavy elements are let
free to adjust, within the observational uncertainties, the  peak is 
completley removed, leaving only a broad bump culminating at 0.6 $R_{\odot}$, 
which presumably is due to another cause ({\it long dashed line}).

\begin{figure}[!htb]
\setlength{\unitlength}{1.0cm}
\begin{picture}(8,4.3)
\includegraphics{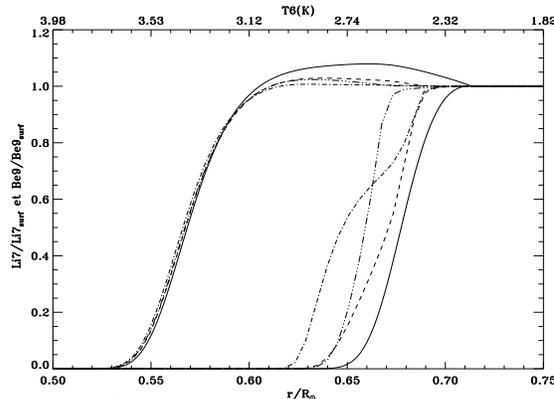}
\end{picture}
\vspace{0.2cm}
\caption[]{\small $^9$Be and $^7$Li radial profile (approximately 
superimposed the temperature scale) for several models: respectively, 
reference ({\it solid line}), coefficient $A_t$ ({\it dash}), coefficient $B_t$ ({\it dash dot}) and coefficient $C_t$ ({\it dash three dots}).}
\end{figure}

The two light elements $^7$Li and $^9$Be are extremely sensitive 
to mixing processes 
occuring in stars because their nuclear burning temperatures are rather low 
(respectively, 2.5 $10^6$, and 3.2 $10^6$ K) [26]. The new 
observational constraints can only be satisfied if those chemical species are 
mixed in a rather thin layer below the convective zone, 
in order to preserve $^9$Be, which is very little depleted according to 
Balachandran and Bell [27].
This is the case with our tachocline model. However, if the mixing had 
proceeded in the
past at the same rate as in the present Sun, $^7$Li would have been
depleted only by a factor $\sim 4$, which is insufficient to account for 
the photospheric lithium abundance ([28] and 
 references therein). 
But the thickness of the tachocline and the strength of mixing have been 
larger in the past, when the Sun was rotating faster. This effect is 
included in the models labeled with the index $t$ (as $B_t$), whose
evolution was calculated with the time-dependent diffusivity of Eq. (4). 

\begin{figure}[!htb]
\setlength{\unitlength}{1.0cm}
\begin{picture}(8,5)
\includegraphics{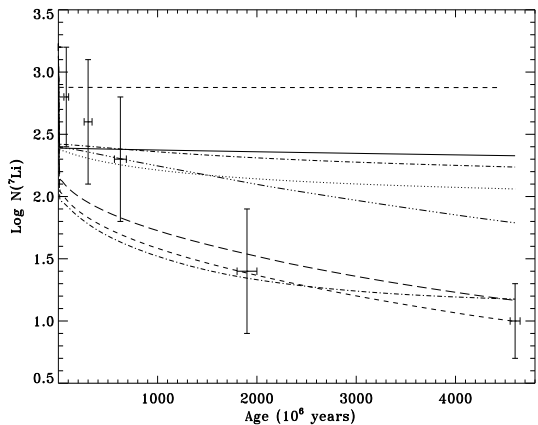}
\end{picture}
 \caption[]{\small Time dependent depletion of $^7$Li for several 
solar 
models: no microscopic diffusion ({\it dash}), with microscopic diffusion ({\it solid line}) and with mixing in the tachocline thickness: coefficient $A$ ({\it dash dot}), $B$ ({\it dash three dots}), then with 
time dependent mixing $A_{t}$ ({\it dots}), $B_{t}$ ({\it long dash}), $C_{t}$ 
({\it thick dash dot}) and $B_{tz}$ (Z$_0=$Z$_0^{ref}=0.01959$) ({\it thick dash}). We superimposed on the theoretical curves the open cluster observations (adapted from Vauclair and Richard [14] and cluster age uncertainties deduced from Lebreton et al. [29]). }
\end{figure}

In Table 1 we give the initial over present ratio 
of $^7$Li and $^9$Be and show in Figure 4 the radial profile of $^7$Li and 
$^9$Be normalized to the surface abundance. We clearly see that the mixing 
process modifies the distribution of lithium but not that of beryllium 
(exception being the flat plateau for the mixed models in comparison with the 
``pure'' diffusive one). With the coefficients $B$ more $^7$Li is burned 
than 
with $A$, and we also see that the time dependence (models with index $t$) 
improves the $^4$He surface abundance as well as the $^7$Li depletion, where a 
value of $\sim$ 100 is obtained without destroying $^9$Be or increasing too 
much 
the $^3$He/$^4$He surface ratio over the past 3 Gyr, as deduced by Geiss and 
Gloeckler [30] from meteorites and solar wind abundance measurements (see 
Table 1). 

In Figure 5 we show the lithium depletion occurring during the Sun's 
evolution for the different models presented, plus a model without any 
diffusion. Clearly, only the diffusive models including mixing in the 
tachocline yield a substantial depletion during main sequence evolution, 
in agreement with the observations (superimposed with their inherent dispersion on the theoretical curves). Note that the strong time dependent mixing with $N=25$ (models $B_{t}$,  $C_{t}$ and  $B_{tz}$) presents a reasonable value of the solar $^7$Li depletion ($\sim$ 100).

However the lithium depletion during the PMS is probably 
overestimated due to the crude spin-down law we have adopted. A more detailed 
analysis of these phases is under study, including metallicity effects and 
more appropriate angular momentum evolution during this phase. 

Our results show the interest to follow together the photospheric abundance 
of the four elements $^3$He, $^4$He, $^7$Li, $^9$Be, and to examine their 
sensitivity to the microscopic, as well as the macroscopic, processes. This 
study encourages the introduction of macroscopic 
 processes in stellar evolution models, and demonstrates the
crucial role of the thin tachocline layer 
 below the convective zone.


\begin{references}
\reference{1.} Corbard, T., L. Blanc-F\'eraud, G. Berthomieu \& J. Provost. 1999.
Non linear regularization for helioseismic inversions. Application for the study of the solar tachocline. Astron. Astrophys. {\bf 344:} 696-708.
\reference{2.} Spiegel, E. A. \& J.-P. Zahn. 1992. The solar tachocline. Astron. Astrophys. {\bf 265:} 106-114.
\reference{3.} Gough, D.O. \& M.E. McIntyre. 1998. Inevitability of a magnetic field in the Sun's radiative interior. Nature. {\bf 394:} 755-757.
\reference{4.} Gilman, P.A. \& P.A. Fox. 1997. Joint instability of latitudinal differential rotation and toroidal magnetic fields below the solar convection zone. Astrophys. J. {\bf 484:} 439-454.
\reference{5.} Chaboyer, B. \& J.-P. Zahn. 1992. Effect of horizontal turbulent diffusion on transport by meridional circulation. Astron. Astrophys. {\bf 253:} 173-177.
\reference{6.} Press, W.H. 1981. Radiative and other effects from internal waves in solar and stellar interiors. Astrophys. J. {\bf 245:} 286-303.
\reference{7.} Schatzman, E. 1993. Transport of angular momentum and diffusion by the action of internal waves. Astron. Astrophys. {\bf 279:} 431-446.
\reference{8.} Choudhuri, A.M., M. Sch\"ussler \& M. Dikpati. 1997. The solar dynamo with meridional circulation. {\bf 319:} 362-362.
\reference{9.} Kosovishev et al. 1997. Structure and rotation of the solar interior: Initial results from the MDI medium-l program. Sol. Phys. {\bf 170:} 43-61.
\reference{10.} Watson, M. 1981. Shear instability of differential rotation in stars. Geophys. Astrophys. Fluid Dynam. {\bf 16:} 285-298.
\reference{11.} Garaud, P. \& D.O. Gough. 1999 (private communication)
\reference{12.} Richard, D. \& J.-P. Zahn. 1999. Turbulence in differentially rotating flows. What can be learned from the Couette-Taylor experiment. Astron. Astrophys. {\bf 347:} 734-738. 
\reference{13.} Zahn, J.-P. 1998. Macroscopic transport. Large-scale advection, turbulent diffusion, wave transport. Space Sc. Rev. {\bf 85:} 79-90.
\reference{14.} Vauclair, S. \& O. Richard. 1998. Consistent solar models including the $^7$Li and $^3$He constraints. in Structure and Dynamics of the Interior of the Sun and Sun-like Stars, S. G. Korzennik \& A. Wilson Eds. ESA SP-418, Vol 1: 427-429. ESA Publication Division, Noordwijk, The Netherlands.
\reference{15.} Brun, A.S., S. Turck-Chi\`eze \& J.-P. Zahn. 1999. Standard solar models in the light of new helioseismic constraints. II. Mixing below the convective zone. Astrophys. J. {\bf 525:} 1032-1041.
\reference{16.} Brun, A.S., S. Turck-Chi\`eze \& J.-P. Zahn. 1998. Macroscopic processes in the solar interior. in Structure and Dynamics of the Interior of the Sun and Sun-like Stars. S. G. Korzennik \& 
A. Wilson Eds. ESA SP-418, Vol 1: 439-443. ESA Publication Division, Noordwijk, The Netherlands. 
\reference{17.} Thompson, M.J., J. Toomre and the GONG Dynamics Inversion Team
1996. Differential rotation and dynamics of the solar interior. Science. {\bf 272:} 1300-1305.
\reference{18.} Gough, D.O. \& T. Sekii. 1997. On the solar tachocline. in IAU 181 Sounding Solar and Stellar Interior (poster volume). J. Provost \& F. X. Schmider Eds: 93-94. Observatoire de la 
C\^ote d'Azur. Nice. France.
\reference{19.} Donahue, R.A., S.H. Saar \& S.L. Baliunas. 1996. A relationship between mean rotation period in lower main-sequence stars and its observed range. Astrophys. J. {\bf 466:} 384-391.
\reference{20.} Skumanich, A. 1972. Time scales for CA II emission decay, rotational braking, and lithium depletion. Astrophys. J. {\bf 171:} 565-567.
\reference{21.} Brun, A.S., S. Turck-Chi\`eze \& P. Morel. 1998. Standard solar models in the light of new helioseismic constraints. I. The solar core. Astrophys. J. {\bf 506:} 913-925.
\reference{22.} Morel, P. 1997. CESAM: A code for stellar evolution calculations. Astron. Astrophys. Sup. {\bf 124:} 597-614.
\reference{23.} Engstler et al. 1992. Test for isotopic dependence of electron screening in fusion reactions. Phys. Lett. B. {\bf 279:} 20-24.
\reference{24.} Basu, S. \& H.M. Antia. 1995. Helium abundance in the solar envelope. Mon. Not. Roy. Astron. Soc. {\bf 276:} 1402-1408.
\reference{25.} Rogers, F.J., J. Swenson \& C. Iglesias. 1996. OPAL equation-of-state tables for astrophysical Applications. Astrophys. J. {\bf 456:} 902-908.
\reference{26.} Baglin, A. \& Y. Lebreton. 1990. Surface abundances of light elements as diagnostic of transport processes in the Sun and solar-type stars. in Inside the Sun. G. Berthomieu \& M. Cribier Eds. Astrophysics and Space Science Library 159: 437-448. Kluwer Academic Publishers. Netherlands.
\reference{27.} Balachandran, S. \& R.A. Bell. 1998. Shallow mixing in the solar photosphere inferred from revised beryllium abundances. Nature. {\bf 392:} 791-793.
\reference{28.} Cayrel, R. 1998. Lithium abundances in low-z stars. Space Sc. Rev. {\bf 84:} 145-154.
\reference{29.} Lebreton Y., A.E. Gomez, J.-C. Mermilliod, M.A.C. Perryman. 1997. 
The age and helium content of the Hyades revisited. in Proceedings of the ESA Symposium 'Hipparcos- Venice '97'. ESA SP-402: 231-236. ESA Publication Division, Noordwijk, The Netherlands.
\reference{30.} Geiss, J. \& G. Gloeckler. 1998. Abundances of deuterium and helium-3 in the protosolar cloud. Space Sc. Rev. {\bf 84:} 239-250. 
\reference
\end{references}
\end{document}